\documentclass[aps,pra,twocolumn,a4paper,superscriptaddress,nofootinbib]{revtex4-1}
\usepackage{color}
\usepackage{graphicx}
\usepackage{bm}
\usepackage{here}

\date{\today}
\begin{document}
\title{A Quantum Model for Coherent Ising Machines:\\ Discrete-time Measurement Feedback Formulation}

\author{Atsushi Yamamura}
\email{atsushi.yamamura@sat.t.u-tokyo.ac.jp}
 \affiliation{Department of Electrical Engineering and Information System, the University of Tokyo, Japan}
 \affiliation{Institute of Industrial Science, The University of Tokyo, Japan}

\author{Kazuyuki Aihara}
\affiliation{Department of Electrical Engineering and Information System, the University of Tokyo, Japan}
 \affiliation{Institute of Industrial Science, The University of Tokyo, Japan}
 \affiliation{Department of Mathematical Informatics, the University of Tokyo, Japan}
\author{Yoshihisa Yamamoto}%
\affiliation{E.L. Ginzton Laboratory, Stanford University, Stanford, CA 94305, USA}

\begin{abstract}
   Recently, the coherent Ising machine (CIM) as a degenerate optical parametric oscillator (DOPO) network has been researched to solve Ising combinatorial optimization problems. We formulate a theoretical model for the CIM with discrete-time measurement feedback processes, and perform numerical simulations for the simplest network, composed of two degenerate optical parametric oscillator pulses with the anti-ferromagnetic mutual coupling. We evaluate the extent to which quantum coherence exists during the optimization process.
\end{abstract}
\maketitle
\section{introduction}
There is significant interest in finding alternatives to modern von-Neuman computers, in particular for solving combinatorial optimization problems in NP-hard and NP-complete classes. Artificial neural networks can implement NP-hard Ising problems \cite{hopfield1985neural} and NP-complete k-SAT problems \cite{ercsey2011optimization}. Other efforts include adiabatic quantum computation \cite{farhi2001quantum}, and quantum annealing \cite{kadowaki1998quantum} with superconducting quantum circuits. However, providing dense connectivity between qubits in such physical systems remains a major challenge in achieving a satisfying efficiency with these alternative approaches \cite{rieffel2015case}.

Recently, many types of Coherent Ising machines (CIM) have been studied to solve Ising-type combinatorial optimization problems \cite{ nigg2017robust, puri2017quantum,goto2017dissipative}. Among these systems, the CIMs using denegerate optical parametric oscillator (DOPO) networks are also being discussed  \cite{wang2013coherent,takata2015quantum,maruo2016truncated}. The first generation of this CIM implements spin-spin coupling through optical delay lines \cite{marandi2014network,roslund2014wavelength,takata201616,inagaki2016large,treps2016quantum}. This type of direct coupling Ising machine can implement $O(N^2)$ spin-spin connections with $N-1$ optical delay lines, which remains still a technical challenge for a large spin size ($N\gg 1$). Recently, the DOPO networks with the measurement feedback circuit were implemented at Stanford University \cite{mcmahon2016fully} and NTT \cite{inagaki2016coherent}. In these machines, the oscillators are coupled indirectly with the  discrete-time quantum measurement and feedback processes, which can generate classical correlations between oscillators. In this paper, we formulate a quantum model with completely positive trace preserving (CPTP) maps of the measurement feedback process for the first time, and numerically evaluate the extent to which quantum coherence exists during the optimization process. This paper is organized as follows. In Section II, we introduce our theoretical model of the DOPO network with the discrete-time measurement feedback processes. In Section III, we present numerical simulation results for a simple DOPO network. Finally, in Section IV we conclude with a brief summary.

\section{The Theoretical Model}
\begin{figure}[h]
  \centering
  \includegraphics[width=9cm]{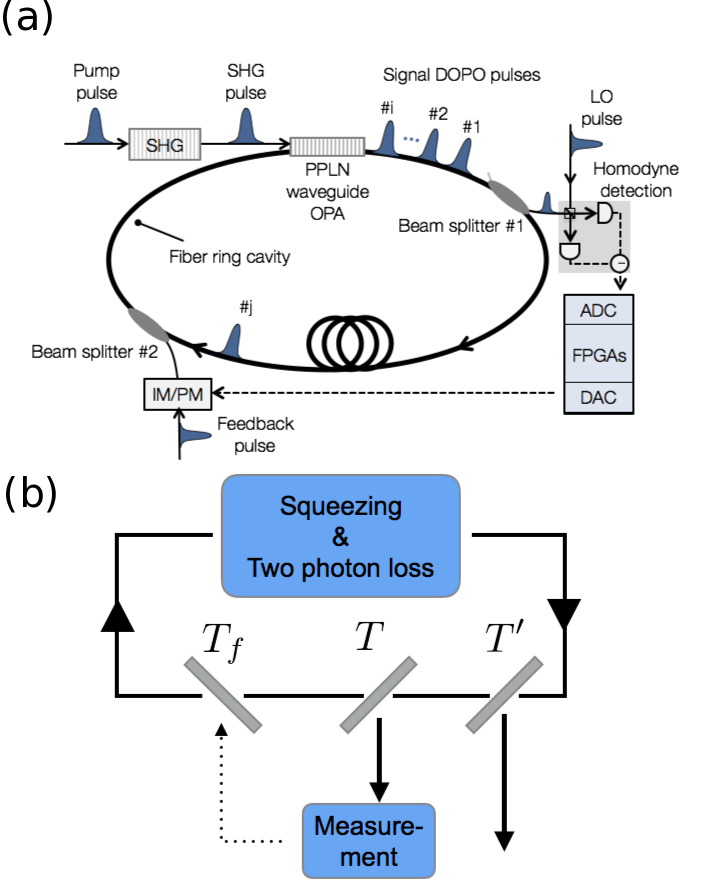}
  \caption{(a) The experimental set up of the coherent Ising machine (CIM) using a degenerate optical parametric oscillator network and discrete-time measurement feedback processes. $N_{pulse}$ pulses are arranged at regular intervals and running around the ring cavity. The amplitude of the feedback pulse injected to the pulse j is calculated by the FPGA from the measurement results of all the other pulses. (b) The corresponding theoretical model. The first beam splitter, with transmittance rate is $T'$, represents the background loss along the ring cavity. }
  \label{fig:model}
\end{figure}

Our theoretical model consists of four components: a PPLN waveguide as a phase sensitive amplifier; two output couplers for simulating the measurement loss and background loss; and a feedback circuit consisting of optical homodyne detectors, an analog-to-digital converter (ADC), a field-programmable gate array (FPGA), a digital-to-analog converter (DAC), and an optical amplitude/phase modulator (Figure \ref{fig:model}). The first output coupler represents all of the background loss in the ring cavity. The signal fields extracted by the second output coupler are used to measure the in-phase amplitudes from the homodyne detectors. In the feedback process, the feedback pulses are generated as coherent states with an average excitation amplitude $x_i=\sum_j J_{ij} \tilde{x_j}$, where $\tilde{x_j}$ is the measurement result for the j th pulse and $J_{ij}$ is the Ising coupling constant. Each signal pulse undergoes these four processes each time it completes one round trip along the ring cavity. To simulate the way in which the states of signal pulses evolve, we calculate the CPTP maps of measurement feedback processes. Then, we numerically simulate the system by expanding the field density operators in terms of the eigenvectors of the in-phase amplitude operator 
$x=(a+a^\dagger)/\sqrt{2}$ \footnote{When we define the in-phase and quadrature-phase amplitude operators as $x=(a+a^\dagger)/\sqrt{2}$ and $ p=(a-a^\dagger)/\sqrt{2}$, they satisfy the commutation relation $[x,p]=i$ and the uncertainty principle $\protect\langle \Delta x^2 \protect\rangle \protect\langle \Delta p^2 \protect\rangle = 1/4$ .}, where $a$ and $a^\dagger$ are the annihilation/creation operators of the signal pulse. 
Because these processes along the ring cavity consists of local operators and classical communications, the states of the signal pulses are not entangled. This is in sharp contrast to the direct optical coupling DOPO network \cite{takata2015quantum,maruo2016truncated}. We calculate conditional density matrices governed by the randomly determined measurement results $x_m$ of the in-phase amplitudes of the signal pulses extracted by the second output coupler.

In the following subsections, we derive the time evolution equation for the density operators and CPTP maps for the four processes in the ring cavity.
For simplicity, we consider a rotating coordinate and ignore the free field Hamiltonian.

\subsection{Phase sensitive amplifier (PSA)}
In the PSA, the signal pulse with an initial state $\rho$, frequency $\omega$, and annihilation operator $a$, interacts with the pump pulse, whose initial state is a coherent state $|\beta\rangle $ ($\beta$ is real) with a frequency $2\omega$. The pump pulse after the interaction is thrown away and in the next round trip, new pump pulse $|\beta\rangle$ is generated.

The amplitude $\beta$ of the pump pulse is very large and the interaction with the signal pulse is weak. 

The Hamiltonian for the PSA is
\begin{eqnarray}
\label{eq:hamiltonian}
\mathcal{H} &=& \frac{1}{2}({e^{2i\omega t}a^\dagger}^2 B + e^{-2i\omega t}a^2 B^\dagger ) .
\end{eqnarray}
Here, $B$ is defined as follows:
\begin{eqnarray}
B = i\sum_{k}  g(\omega_k) e^{-i\omega_k t} b(\omega_k),
\end{eqnarray}
where $ b(\omega_k)$ is the annihilation operator of the pump field of frequency $\omega_k$, and $ g(\omega_k)$ is a parametric coupling constant.
Initially, the pump field at $2\omega$ is in a coherent state $|\beta \rangle$ and all the other modes are assumed to be in vacuum state $|0\rangle$. When we displace the pump field as follows
$$
b(2\omega) \rightarrow b(2\omega) + \beta,
$$
the initial state of the pump field is displaced to a vacuum state.
In this case, the Hamiltonian (\ref{eq:hamiltonian}) is rewritten as
\begin{eqnarray}
\mathcal{H} &=& g(2\omega) \frac{i}{2}({a^\dagger}^2 \beta - a^2 \beta^* ) +  \frac{1}{2}({e^{2i\omega t}a^\dagger}^2 B + e^{-2i\omega t}a^2 B^\dagger )  \nonumber \\
&=&: \mathcal{H}_s + \mathcal{H}_{int}.
\end{eqnarray}

We shall interpret that the first term $\mathcal{H}_s$ represents the linear phase sensitive amplification/ deamplification (squeezing effect) imposed by the pump, while the second term represents up conversion process, which results in an effective two photon loss of the signal mode. The first part of the Hamiltonian can be absorbed as an internal Hamiltonian of the signal. Then, the second part is expressed as
\begin{eqnarray}
\mathcal{H}^I_{int} &=& \frac{1}{2}({e^{2i\omega t}a^{I\dagger}}^2(t) B + e^{-2i\omega t}{a^I}^2(t) B^{\dagger} ),
\end{eqnarray}
in the interaction picture.
We take the Born-Markov and the rotating wave approximations to eliminate the Hilbert space of the pump pulse. Then, we can find the master equation corresponding to the interaction Hamiltonian as follows \cite{breuer2002theory} :
\begin{eqnarray}
\frac{d\rho^I}{dt} &=& \sum_k  \left[ (\Gamma_{k} + \Gamma_{k}^* ) {a^I}^2(t) \rho {a^{I\dagger}}^2(t) \right. \nonumber \\ && \left.
- \Gamma_{k} {a^I}^2(t)  {a^{I\dagger}}^2(t)  \rho - \Gamma_{k}^* \rho {a^{I\dagger}}^2(t)  {a^I}^2(t)  \right],
\end{eqnarray}

where
\begin{eqnarray}
\Gamma_{k} &=& \int_0^\infty ds e^{i(2\omega-\omega_k)s} g(\omega_k)^2 \langle b(\omega_k)b^\dagger(\omega_k)\rangle \nonumber \\
&=& \frac{g(2\omega)^2}{2}\delta(2\omega - \omega_k) \langle b(\omega_k)b^\dagger(\omega_k)\rangle \nonumber \\
 &=& \frac{g(2\omega)^2}{2} \delta(2\omega - \omega_k).
\end{eqnarray}
When we go back to the Schrodinger picture, we have
\begin{eqnarray}
\frac{d\rho}{dt} &=& \frac{g(2\omega)\beta}{2} \left[({a^\dagger}^2 - a^2), \rho \right] \nonumber \\ 
&+& \frac{g(2\omega)^2}{2} \left[2 {a}^2 \rho {a^{\dagger}}^2
- {a^{\dagger}}^2 {a}^2  \rho - \rho {a^{\dagger}}^2  {a}^2  \right].
\label{eq:schro}
\end{eqnarray}

The first term in Eq.(\ref{eq:schro}) represents a standard unitary (squeezing) process, while the second term is a Lindblad form representing the two photon loss process associated with the parametric pump photon generation.
We defined the squeezing rate $S=g(2\omega)\beta t$ and the two photon loss rate $L=g(2\omega)^2 t$, where $t$ is the time duration of this interaction in the PSA. The linear power-gain $G$ can be represented as $G = \exp(2S) = \exp(2g(2\omega)\beta t)$.
One of the important assumption leading to Eq.(7) is that the gain saturation is relatively weak, i.e. the signal pulse intensity grows and depletes the pump power only slightly, instead of decaying to zero, as is the case for a traveling-wave PSA with strong signal-pump interaction \cite{hamerly2016topological}. The other important assumption is that the pump field is dissipated into external reservoirs each time of PSA and a fresh coherent state $|\beta\rangle$ is always prepared as a new pump field for the next round of PSA.
The equation for our simulation is obtained by expanding the density operator $\rho$ in terms of the in-phase amplitude eigenstates $|x\rangle$ as follows:
\begin{eqnarray}
\frac{d}{dt} \langle x|\rho|x' \rangle &=& g(2\omega)\beta \left( -z\partial_z - w \partial_{w} - 1 \right) \langle x|\rho|x' \rangle \nonumber \\ 
&+& \frac{g(2\omega)^2}{8} \left( -z^2w^2 + 3(z^2+w^2) \right. \nonumber \\ && \left. + (z^2-w^2+8)(z\partial_z + w \partial_w) + 4(z^2-1)\partial_z^2  \right. \nonumber \\ && \left.+4(w^2-1)\partial_w^2 + 4(z\partial_z - w\partial_w)(\partial_z^2 - \partial_w^2)  \right. \nonumber \\ && \left. - 16\partial_z^2 \partial_w^2   \right) \langle x|\rho|x' \rangle,
\end{eqnarray}
where $z=x+x'$ and $w=x-x'$.

All coefficients which appears in the equation above are real-valued because annihilation and creation operators, which appear in the equation (\ref{eq:schro}), are summation of in-phase amplitude operator $x$ and its derivative $\partial_x$ with real coefficients, which allows $\langle x|\rho|x' \rangle$ are real-valued in CIM.

\subsection{Output couplers and homodyne detectors}
A portion of the signal pulse is extracted from the ring cavity by the two beam splitters. At the first output coupler, the extracted signal-field is simply dissipated in external reservoirs, which represents the background loss in the ring cavity. At the second output coupler, the in-phase amplitude $x=(a+a^\dagger)/\sqrt{2}$ of the extracted field is projectively measured by the homodyne detectors. We define the transmittance of the first and second splitters as $T'=\sin^2 \theta'$ and $T=\sin^2 \theta$.  When the signal pulse goes into the beam splitter, it is combined with the incident vacuum state from the external environments. Thus, the measurement performed by the homodyne detectors has a finite measurement error, which stems from the vacuum fluctuation. We define the annihilation operators of the signal and the vaccum field as $a$ and $a_{vac}$. Then, the output field annihilation operators can be written in terms of the unitary operator $U$ of the beam splitter with a parameter $\theta$ as follows:

\begin{equation}
U^\dagger a U = \sin\theta a_{vac} +\cos \theta a,
\end{equation}
\begin{equation}
U^\dagger a_{vac} U = \sin\theta a +\cos \theta a_{vac}.
\end{equation}

From these equations, the Kraus operator corresponding to the measured value of $x_m$ can be expressed by
\begin{eqnarray}
M_{x_m} &=& \langle x_m|U|0\rangle  \nonumber \\ &=& \int dx_i dx_f \pi^{-1/4} \delta(x_i- (\cos\theta x_f+\sin\theta x_m))  \nonumber \\ && \times \exp(-\frac{1}{2}(-\sin\theta x_f+\cos\theta x_m)^2) |x_f\rangle \langle x_i|. \nonumber \\
\end{eqnarray}
To calculate the conditional density matrix for the post-measurement state, we generate a random number and determine a measured value $x_m$ with the probability of $Tr(M_{x_m} \rho M_{x_m}^\dagger)$.
For the operation for the first output coupler, we use the same Kraus operators $\{M_{x_m}\}$ and ensemble many conditional density matrices governed by the probabilistically determined $x_m$.

\subsection{Feedback process}
In the feedback injection process, the signal pulse and feedback pulse, the latter of which is prepared in a coherent state $|\alpha\rangle $, are combined with a third beam splitter. The transmittance  rate of the third beam splitter, defined as $T_f=\sin^2\theta_f$, is very high ($T_f \approx 1$). In this parameter region, the quantum flucutation due to the injected coherent state is much smaller than the flutucation in the signal pulse, and this feedback process can be described with a simple unitary displacement operator   $D(\alpha\theta_f) = \exp(\alpha \theta_f a^{\dagger} - \alpha^* \theta_f a)$.
In the Heisenberg picture, the in-phase amplitude operator $x$ will be translated as $D(\alpha\theta_f)x D^\dagger(\alpha\theta_f) = x+ \alpha\theta_f /\sqrt{2}$.

The amplitude $\alpha$ of the feedback pulse  is determined by the measured values of the homodyne detectors. We define the feedback rate $R$ as the ratio of $\alpha\theta_f /\sqrt{2}$ to the in-phase amplitude $x$ of the signal pulse, estimated using the value $x_m$ measured by the homodyne detectors.

\subsection{Summary of the modeling}
In summary, the signal pulses experience four processes described above during each loop around the ring cavity. The system is described by five physical parameters, the gain $G$ (or squeezing rate $S$), two photon loss rate $L$, background loss rate $1-T'$
, measurement loss rate $1-T$, and feedback rate $R$.
The total net linear amplitude-gain, before the gain saturation is switchwed on, during one round trip of the cavity is $G_{tot}=\sqrt{G\times T \times T'} = \exp(g(2\omega)\beta t) \sin\theta \sin\theta'$, where we assume that $T_f \approx 1$.\\

\begin{figure}[h]
  \centering
  \includegraphics[width=9cm]{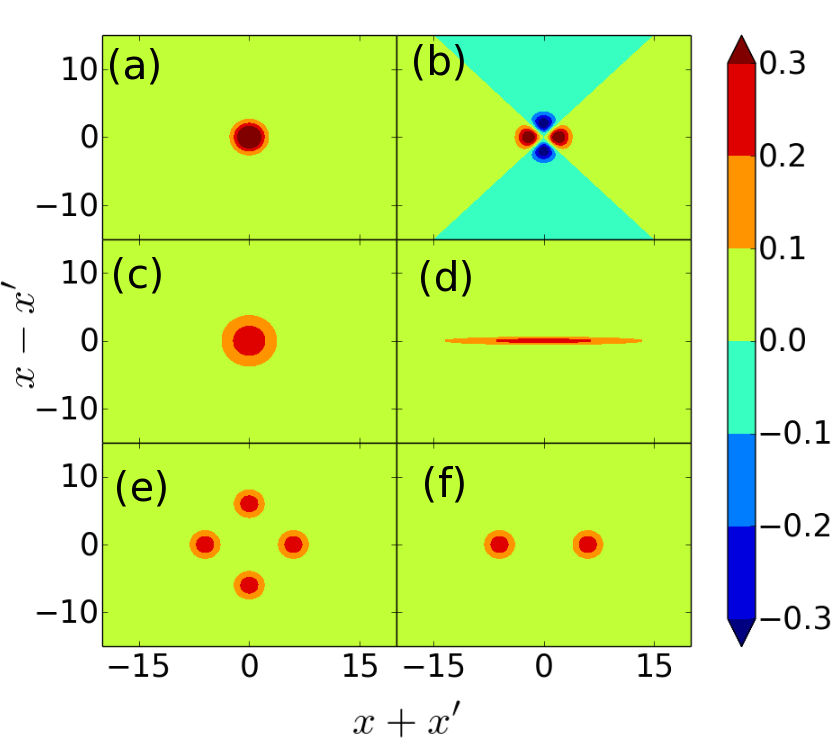}
  \caption{Contour maps of the density matrices $\langle x| \rho |x'\rangle$ of typical quantum states as functions of $x+x'$ and $x-x'$. $\langle x| \rho |x'\rangle$ are real fuctions in the cases of these states. (a) a vacuum state $|0\rangle$ (b) a one photon state $|1\rangle$ (c) an anti-squeezed vacuum state (d) a thermal state ($\langle n\rangle = 2$) (e) a cat state of coherent states $\frac{1}{\sqrt{2}}(|\alpha \rangle + |-\alpha \rangle)$ (f) a classical ensemble of coherent states  $\frac{1}{2}(|\alpha\rangle \langle\alpha| + |-\alpha \rangle \langle - \alpha|)$.}
  \label{fig:examples}
\end{figure}

For numerical simulations, we expand the conditional density matrices of pulses in terms of $x$-eigenvectors and calculate the elements of the density matrices. Note that the elements of density matrices $\langle x| \rho |x'\rangle$ are real numbers in this system. In this paper, we will visualize density matrices as functions of $x+x'$ and $x-x'$. While this function is equivalent to the Wigner function because Wigner function can be obtained by the Frourier transformation of $\langle x| \rho |x'\rangle$ along the $x-x'$ axis, it is easier to see the quantum coherence than Wigner function as shown in Figure \ref{fig:examples}.

 Figure \ref{fig:examples} shows the contour maps of the functions $\langle x| \rho |x'\rangle$ corresponding to typical quantum states. The line of $x-x'=0$ is the diagonal line of a density matrix and represents the probability distribution on $x$, while the line of $x+x'=0$ represents quantum coherence between the element of $|x \rangle$ and the one of $|-x\rangle$. The vacuum state $|0\rangle$ and a one photon state $|1\rangle$ can be represented with a simple gaussian function whose variance is $\langle \Delta x^2 \rangle = 0.5$ and a phase reversed Hermite Gaussian function (Figure \ref{fig:examples}(a),(b)). When it is anti-squeezed along the $x$-axis, it becomes broader gaussian distribution (Figure \ref{fig:examples}(c)). On the other hand, a thermal state is a gaussian whose width is large along the $x+x'$ axis but small along the $x-x'$ axis (Figure \ref{fig:examples}(d)). This clearly shows the low qunatum coherence in the thermal state. While non-zero values appear along the vertical line of $x+x'=0$ when two coherent states $|\alpha\rangle$ and $|-\alpha\rangle$ are superposed with quantum coherence, they disappear when the two coherent states are classically ensembled (Figure \ref{fig:examples}(e)(f)). These states shown in Figure 2 and states in our CIM model don’t have imaginary elements, so the functions can be described with contour maps of real values.

\section{Numerical Simulation Results}

We simulated the time evolution of the simplest DOPO network, consisting of two oscillators interacting with out-of-phase coupling (anti-ferromagnetic coupling). We consider that a $0$ phase pulse is an up spin, and that a $\pi$ phase pulse is a down spin. In this case, the two degenerate ground states are $0$ phase-$\pi$ phase (up-down $|\uparrow \downarrow \rangle$) and $\pi$ phase-$0$ phase (down-up $| \downarrow \uparrow \rangle$) states.

\subsection{Time evolution of typical conditional density matrices}
\begin{figure}[h]
  \centering
  \includegraphics[width=8cm]{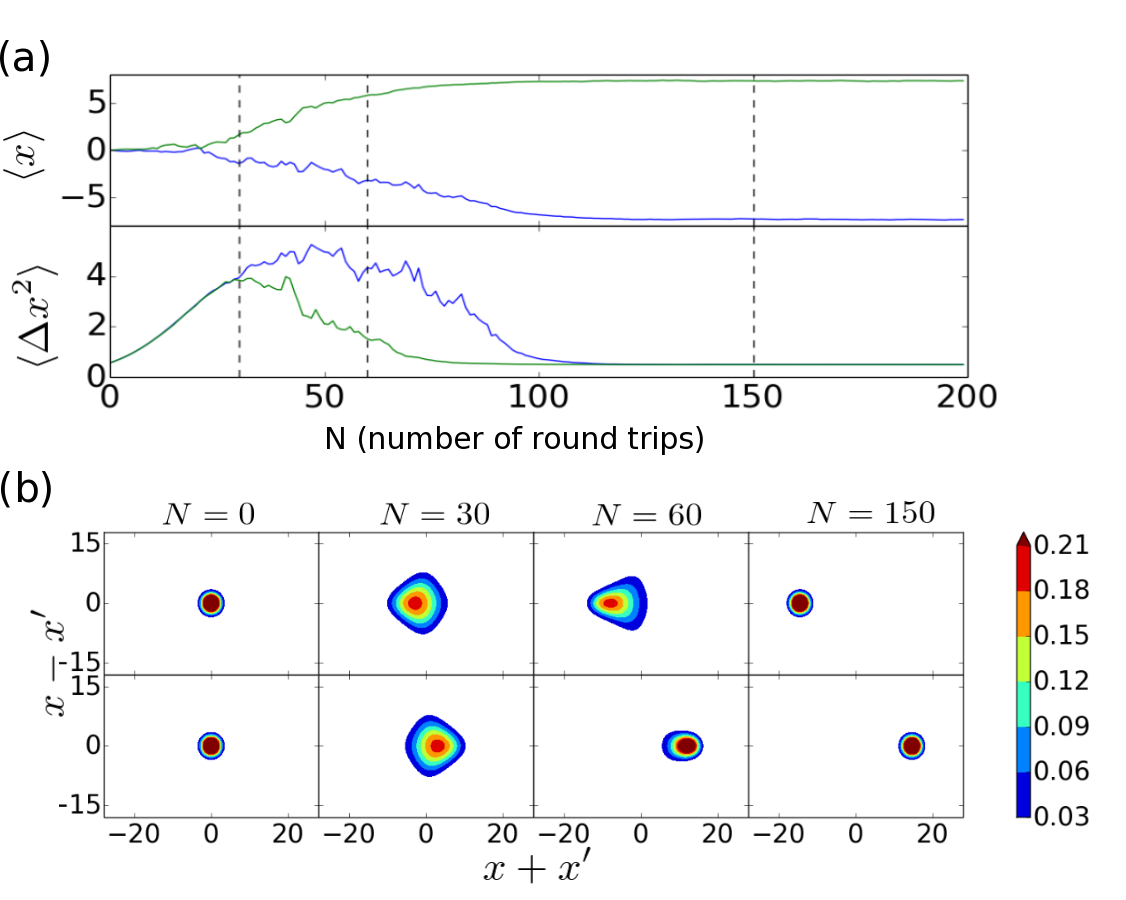}
  \caption{Time evolution of typical conditional density matrices of the two oscillators governed by the measurement results $x_m$ with $T=0.99$. Other parameters are shown in Table \ref{table:data_type}. (a) The evolution of $\langle x \rangle$ and $\langle \Delta x^2 \rangle$. (b) The contour maps of the typical conditional density matrices $<x|\rho|x'>$ in front of the PSA plotted on the coordinates $x-x'$ and $x + x'$. In this system, all elements of the density matrices are real numbers.}
  \label{fig:T099}
\end{figure}

In this subsection, we assume that the background loss in the cavity is zero ($T' =1$), and the transmittance of the output coupler is $T=0.99$. Thus, in this case, the ratio of power extracted from the cavity for the measurement is 0.01. Other numerical parameters are given in Table \ref{table:data_type}.
Figure \ref{fig:T099} illustrates the time-development of typical conditional density matrices of signal pulses governed by the sequence of measured values $x_m$.
The initial states of the two pulses are vacuum states ( with the number of round trips $N=0$ in Figure \ref{fig:T099} (b)). At that time, $\langle x\rangle =0$ and $\langle \Delta x^2\rangle  = 0.5$. The optimization process consists of three stages. In the first stage, the in-phase amplitudes of two pulses are anti-squeezed by the phase sensitive amplifier, and $\langle \Delta x^2\rangle $ become larger ($N=30$ in Figure \ref{fig:T099} (b) ). Note that both the diagonal x-distribution along the horizontal axis and the off-diagonal quantum coherence plotted along the vertical axis become broad. In the second stage, because of the gain saturation and feedback processes, the expectation value $\langle x\rangle $ moves to either negative or positive value ($N=60$ in Figure \ref{fig:T099} (b)). The gain saturation and the linear photon loss are responsible for the spontaneous symmetry breaking of DOPO, while the feedback process makes the system to select an anti-ferromagnetic order instead of a ferromagnetic order. Finally, in the third stage the state becomes close to the highly excited coherent state and $\langle \Delta x^2\rangle $ is reduced to $0.5$ ($N=150$ in Figure \ref{fig:T099} (b) ). At this stage, the DOPO state is already in a classical level and the optimization process of the CIM is completed.
\begin{table}[hbtp]
  \caption{The parameters for numerical simulation for the coherent Ising machine with two DOPOs.}
  \label{table:data_type}
  \centering
  \begin{tabular}{lcc}
    \hline
     physical meaning & name  &  value  \\
    \hline \hline
    Net Gain in One loop  & $G_{tot}$  &   1.05  \\
    Background Loss Rate & $1-T'$  & 0 \\
    Feedback Rate  & $R$ & 0.005 \\
    Two Photon Loss Rate & $L$  & 0.002 \\
    \hline
  \end{tabular}
\end{table}

\begin{figure*}[htbp]
  \centering
  \includegraphics[width=14cm]{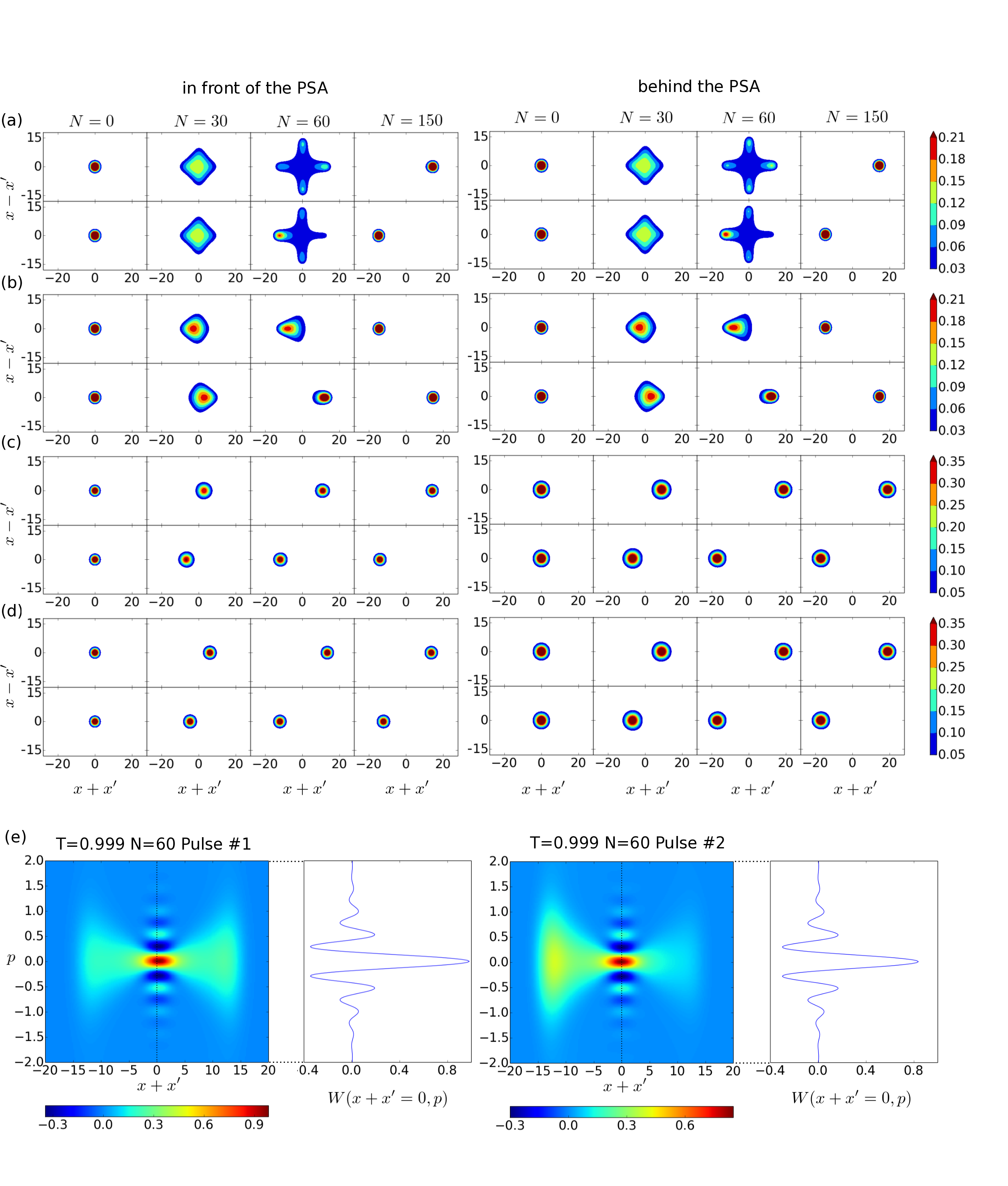}
  \caption{The contour maps of typical conditional density matrices $<x|\rho|x'>$ of two oscillators governed by the measurement results $x_m$  with (a) $T=0.999$, (b) $T=0.99$, (c) $T=0.9$, and (d) $T=0.5$ at round trips of $N=0,30,60,150$  in front of and behind the PSA. Other numerical parameters are shown in Table \ref{table:data_type}. In this system, all elements of the density matrices are real numbers. The figure (e) represents the Wigner function $W(x+x',p)= \frac{1}{\pi} \int^\infty_{-\infty} \langle x| \rho |x'\rangle e^{2ip(x-x')} d(x-x')$ of the density matrices at N=60 in the figure \ref{fig:variousT} (a).}
  \label{fig:variousT}
\end{figure*}

\subsection{Various measurement strengths}
Next, we present the simulation results for various measurement strengths 1-$T$ under a condition of no background loss, $T'=1$. The parameters for the numerical simulations are shown in Table \ref{table:data_type}. Figure \ref{fig:variousT} shows typical conditional density matrices $<x|\rho|x'>$  governed by the sequence of measurement results $x_m$.
It can be seen that the anti-squeezing effect at the early period of the optimization process is more significant when the transmittance of the output coupler is larger or the measurement strength is weaker so that the wavepacket reduction is not so significant.  When $T=0.9$ or $T=0.5$, the states undergo mild anti-squeezing and they are quickly displaced.  On the other hand, when $T=0.999$, the states maintains the quantum coherence between the macroscopically separated "up state" and "down state" ($N=60$ in Figure \ref{fig:variousT} (a)). Here, the probability distributions are not localized but the centers of the two wavepackets are negatively correlated. As shown in Figure \ref{fig:variousT}(e), the Wigner function features an oscillatory behavior with negative amplitudes that manifests the quantum interference effect between the macroscopically separated "up-state" and "down-state".

\begin{figure}[h]
  \centering
  \includegraphics[width=9cm]{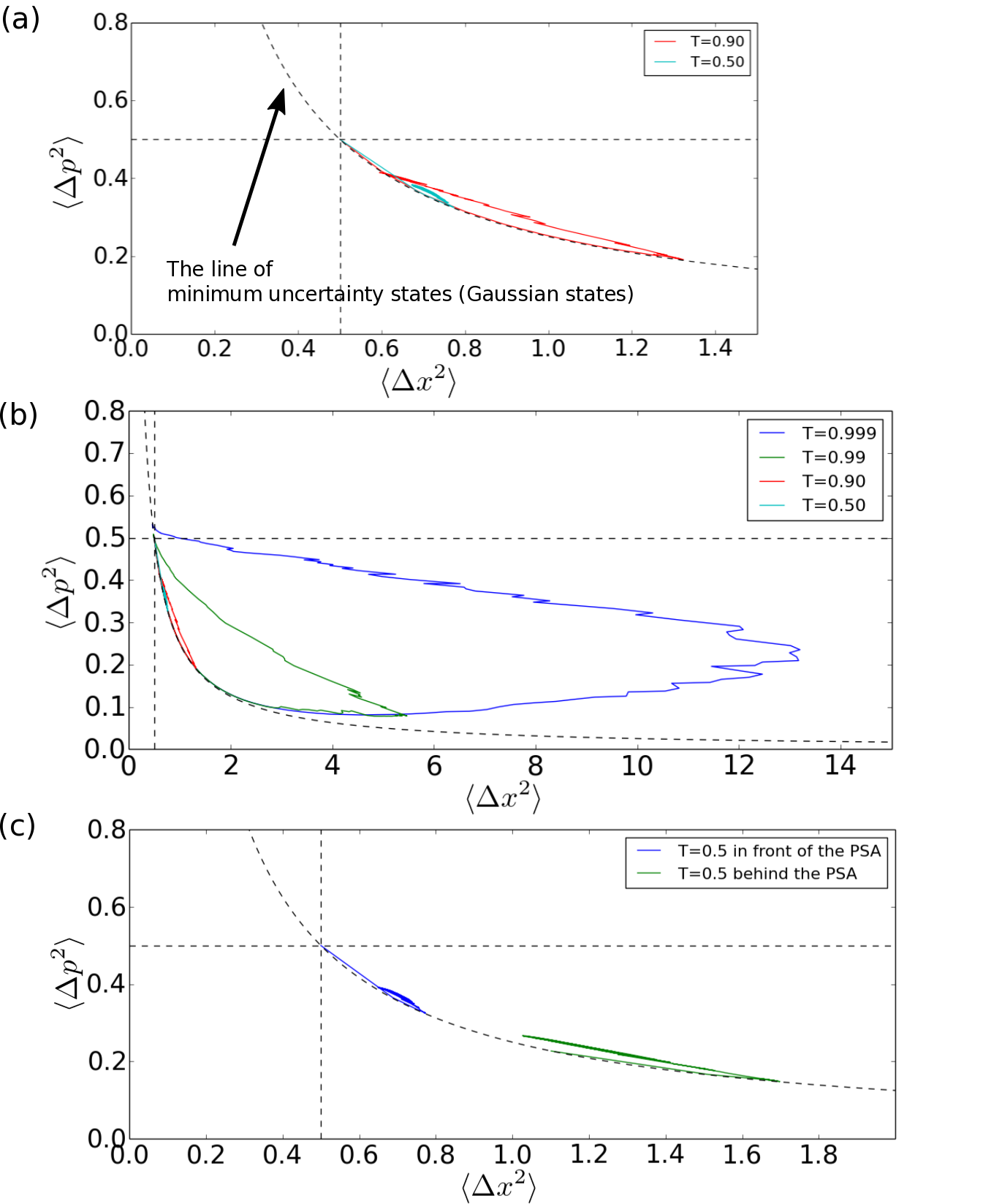}
  \caption{$\langle \Delta x^2\rangle $ vs. $\langle\Delta p^2\rangle $ for a typical conditional density matrix of a signal pulse governed by the sequence of randomly determined measurement results $x_m$, with various measurement strengths. (a) $\langle \Delta x^2\rangle $ vs. $\langle\Delta p^2\rangle $ for $T=0.5$ and $0.9$. (b) $\langle \Delta x^2\rangle $ vs. $\langle\Delta p^2\rangle $ for $T=0.5, 0.9, 0.99$, and $0.999$.  (c) $\langle \Delta x^2\rangle $ vs. $\langle\Delta p^2\rangle $ in front of and behind the PSA in the case of $T=0.5$.
  The dashed curves represent the line of the minimum uncertainty product $\langle \Delta x^2\rangle \langle \Delta p^2\rangle = 1/4$. The initial state of the signal pulse is the vacuum state ($\langle \Delta x^2\rangle =\langle \Delta p^2\rangle =0.5$). }
  \label{fig:coherence}
\end{figure}

Figure \ref{fig:coherence} illustrates the relationship between $\langle \Delta x^2\rangle $ and $\langle \Delta p^2\rangle $ for a typical conditional density matrix and the full density matrix of a one signal pulse for various values of transmittance rate $T$. Here, $x,p$ are the in-phase and quadrature-phase amplitudes, respectively, defined by $x=\frac{a+a^\dagger}{\sqrt{2}},ip=\frac{a-a^\dagger}{\sqrt{2}}$. Initially at N=0, the state satisfies $\langle \Delta x^2\rangle =\langle \Delta p^2\rangle =0.5$. As the signal pulses complete many round trips around the ring cavity, the value of $\langle \Delta x^2 \rangle $ for a conditional density matrix first becomes larger, and then decreases to $\langle \Delta x^2\rangle = 0.5$ at well above DOPO threshold. These lines form loops. The dashed curve represents the minimum uncertainty product. The Heisenberg uncertainty principle dictates that $\langle \Delta x^2\rangle \langle \Delta p^2\rangle \geq 1/4$. Because the state is squeezed vacuum state in the early stage of the optimization process, it lies on this dashed curve when $\langle \Delta x^2\rangle $ is relatively small. 
The results depicted in Figure \ref{fig:coherence} clearly demonstrate the two facts. First, when the measurement strength is weaker, the states are more anti-squeezed and more quantum coherence is present between the "up state" (the region $x >0$) and the "down state" (the region $x<0$).
Second, as $T$ becomes smaller or the measurement strength increases, the conditional density matrix becomes close to a squeezed coherent state, because the density matrix moves almost onto the curve of the minimum uncertainty product. In this case, the density matrices will be efficiently simulatable with the displaced squeezing basis method discussed in \cite{tezak2017low}.

The degree of squeezing is more than 3dB in Figure \ref{fig:coherence} . It is well-known that a continuous wave (CW) pumped DOPO features only 3dB squeezing inside a cavity. However, such a limit does not exist for a pulsed DOPO. The numerical results shown in Figure \ref{fig:coherence}(a)(b) are the uncertainty product at the input to the PSA, while that at the output of the PSA is shown in Figure \ref{fig:coherence}(c).

\subsection{Probability of success and the effect of background loss}
After the many round trips of signal pulses, we projectively measure the in-phase amplitude of the pulses. If the amplitude is positive $x>0$, we treat it as a up-spin state, and otherwise, a down-spin state. Thus, in the case of two signals with out-of-phase coulping, the optimization by CIM is successful if the projectively measured in-phase amplitude of a signal is positive and the one of the other signal is negative. Thus, the definition of the success probability is $\int_{x_1x_2 <0 } \langle x_1 |\langle x_2 | \rho |  x_1 \rangle |  x_2 \rangle dx_1 dx_2 $, where $x_1$ and $x_2$ denote the in-phase amplitudes of the two signal pulses and $\rho$ is the full density matrix for the two signal DOPO pulses.

\begin{figure*}[t]
  \centering
  \includegraphics[width=12cm]{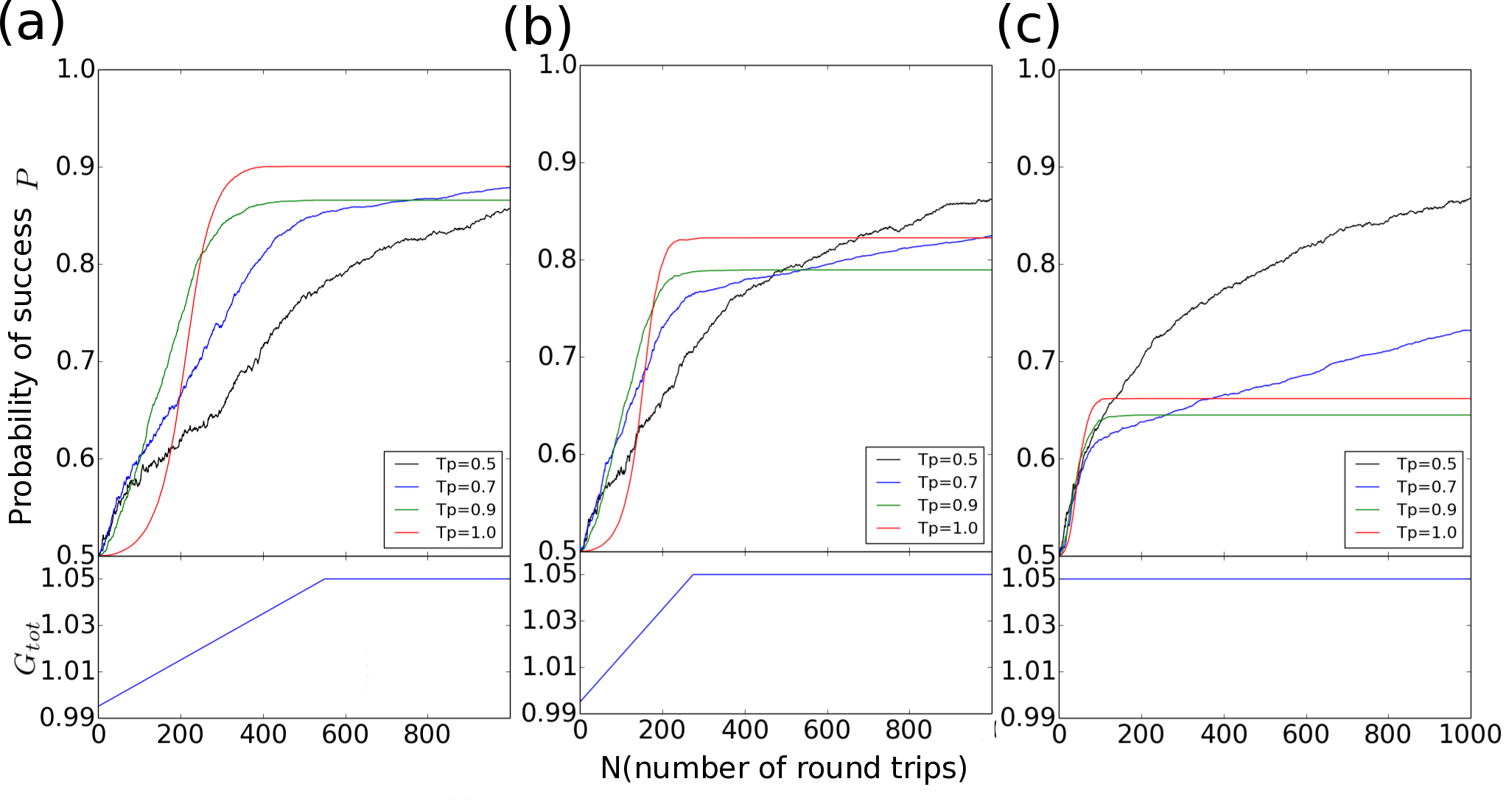}
  \caption{The probability of success $P$ for various values of background loss rate $1-T'$ with three pump schedule for net linear gain. (a) slowly increasing pump rate (b) rapidly increasing pump rate (c) constant pump rate.
   The definition of the success rate is $\int_{x_1x_2 <0 } \langle x_1 |\langle x_2 | \rho |  x_1 \rangle |  x_2 \rangle dx_1 dx_2 $, where $x_1$ and $x_2$ are the in-phase amplitudes for the two oscillators. These values are calculated from the averages of the probabilities of 3000 conditional density matrices for $T'=1.0, 0.9, 0.7,$ and $0.5$. Other parameters for the numerical simulation are shown in Table \ref{table:data_type2}.
  }
  \label{fig:suc}
\end{figure*}

We produced many conditional density matrices, in order to calculate how the probability of success $P$ of the optimization depends on the background loss rate $1-T'$, with the numerical parameters presented in Table \ref{table:data_type2} and three different pump schedules of net linear amplitude-gain $G_{tot}$.
 The results are shown in Figure \ref{fig:suc}.

  The initial success rate for the two vacuum states is $0.5$ and as $N$ increases it becomes higher. To see the dependance of the probability of success on the background loss rate and the time schedule of the net linear gain, we set the feedback rate $R$ not to be sufficiently strong and $P$ does not reach 100\%. Of course, when $R$ is sufficiently strong, $P$ reaches to 100\%. In the case of a low background loss ($T'=1.0$ and $0.9$), the initial increasing rate of $P$ is smaller than the case of a larger background loss. When the loss rate is low, the state is largely anti-squeezed, and this makes the signal-to-noise ratio of measurement is poor at the early stage, as shown in Figure \ref{fig:variousT}.
 Thus, the in-phase amplitudes of the feedback pulses generated from the measured values suffer from small signal-to-noise ratio, and this leads to a lower increase rate for the probability of success at the early stage.

However, once $P$ starts to increase, it suddenly goes up and reaches to the final constant value. The final value greatly depends on the time shedule of the net gain $G_{tot}$.
 When $G_{tot}$ increases rapidly, the final probability of success for the low-loss case becomes lower than that in the case of slowly increasing $G_{tot}$. This is because the small fluctuation leads to the states of the DOPO pulses being easily trapped into the potential wells of DOPO (up-spin state or down-spin state) before the correlation between the two pulses permanently forms. On the other hand, when the background loss is large or $T'$ is small ($T'=0.5$ and $0.7$), the state fluctuates strongly and can move from the up (down) state to the down (up) state even after $G_{tot}$ becomes larger than one (above the threshold)\cite{kinsler1991quantum} . Thus, the success rate continues to increase also in the later stage of the optimization process.

\begin{table}[hbtp]
  \caption{Parameters for the numerically simulation in Figure \ref{fig:suc}.}
  \label{table:data_type2}
  \centering
  \begin{tabular}{lcc}
    \hline
     physical meaning & name  &  value  \\
    \hline \hline
    Measurement Strengt{}h  & $T$  &   0.99  \\
    Feedback Rate  & R & 0.005 \\
    Two Photon Loss Rate & L  & 0.002 \\
    \hline
  \end{tabular}
\end{table}

\section{Conclusions}
We developed the quantum theory for the coherent Ising machines with the discrete-time measurement feedback process. In this optical system, signal pulses in a ring cavity are weakly measured and  the mutual coupling via feedback pulses allows them to interact with each other. The CIM can detect the ground state of the Ising Hamiltonian with a certain probability. In this paper, we showed the simulation results of the simplest DOPO network, which is composed of two anti-ferromagnetically coupled signal pulses.

We showed that there are three stages in the optimization process: anti-squeezing of an initial vacuum state by phase-sensitive amplification, spontaneous symmetry breaking by gain saturation and linear photon loss with dispacement by the feedback process, and excess noise supression to approach the final coherent states. 

We discussed quantum coherence during the optimization process. When the loss rate is small, the signal pulses undergo strong anti-squeezing along the in-phase amplitude, and quantum coherence between different amplitudes is maintained. On the other hand, in the case of a large loss rate, the state is close to the squeezed coherent states, and close to the Heisenberg limit $\langle \Delta x^2\rangle \langle \Delta p^2 \rangle \approx 1/4$ during the optimization process.

We also calculated the success probabilities for different values of background loss rates $1-T'$. In the case of a small background loss, the higher success probability is obtained with a slower increase in the net gain. On the other hand, in the case of a large background loss, the success probability is not as sensitive to the net gain increase schedule.

\section*{Acknowledgement}
This research was funded by the Impulsing Paradigm Change through Disruptive Technologies (ImPACT) Program of the Council of Science, Technology and Innovation (Cabinet Office, Government of Japan). The authors wish to thank Taime Shoji, Ryan Harmerly, Peter Drummond and Hideo Mabuchi for their critical discussions.
\bibliographystyle{apsrev4-1}
\bibliography{manu_ab}
\end{document}